\documentclass[conference]{IEEEtran}
\IEEEoverridecommandlockouts
\usepackage{cite}
\usepackage{amsmath,amssymb,amsfonts}
\usepackage{algorithm2e}
\usepackage{graphicx}
\usepackage{subfigure}
\usepackage{textcomp}
\usepackage{float}

\def\BibTeX{{\rm B\kern-.05em{\sc i\kern-.025em b}\kern-.08em
    T\kern-.1667em\lower.7ex\hbox{E}\kern-.125emX}}

\begin{document}

\title{Triangle-mapping Analysis on Spatial Competition and Cooperation of Chinese Cities \\
\thanks{This work is supported by the National Natural Science Initiative (Nos. 11622538, 61673150, 61673151, 11205040) and the research start-up fund of Hangzhou Normal University (No. 2015QDL005)). L.L acknowledge the Zhejiang Provincial Natural Science Foundation of China (No. LR16A050001). Correspondence should be addressed to X.-P.H (xp@hznu.edu.cn).}
\author{\IEEEauthorblockN{
Pan Liu\IEEEauthorrefmark{1},
Xiao-Pu Han\IEEEauthorrefmark{1},
and
Linyuan L\"{u}\IEEEauthorrefmark{1},\IEEEauthorrefmark{2}
}
\IEEEauthorblockA{\IEEEauthorrefmark{1}
Alibaba Research Center for Complexity Sciences, Hangzhou Normal University, Hangzhou 311121, China\\
Email: xp@hznu.edu.cn}
\IEEEauthorblockA{\IEEEauthorrefmark{2}Institute of Fundamental and Frontier Sciences\\
University of Electronic Science and Technology of China, Chengdu 610054, China\\
Email: linyuan.lv@gmail.com}
}}

%Email: wuleilei868@gmail.com; linyuan.lv@gmail.com}
%\IEEEauthorblockA{\IEEEauthorrefmark{2}School of Software, Jiangxi Normal University, Nanchang 330022, China\\
%Email: dxf@jxnu.edu.cn}
%\IEEEauthorblockA{\IEEEauthorrefmark{3}Computational Social Science, ETH Zurich, Zurich CH-8092, Switzerland\\
%Email: xren@ethz.ch}
%\IEEEauthorblockA{\IEEEauthorrefmark{4}Alibaba Research Center for Complexity Sciences, Hangzhou Normal University, Hangzhou 311121, China\\
%Email: chunxiaojia@163.com}
%\IEEEauthorblockA{\IEEEauthorrefmark{5}College of Information Technology, Jiangxi University of Finance and Economics, Nanchang 330013, China\\
%Email: zhong.ys@163.com}}}
\maketitle
\begin{abstract}
 In this paper, we empirically analyze the spatial distribution of Chinese cities using a method based on triangle transition. This method uses a regular triangle mapping from the observed cities and its three neighboring cities to analyze their distribution of mapping positions. We find that obvious center-gathering tendency for the relationship between cities and its nearest three cities, indicating the spatial competition between cities. Moreover, we observed the competitive trends between neighboring cities with similar economic volume, and the remarkable cooperative tendency between neighboring cities with large difference on economy. The threshold of the ratio of the two cities' economic volume on the transition from competition to cooperation is about $1.2$. These findings are helpful in the understanding of the cities economic relationship, especially in the study of competition and cooperation between cities.
\end{abstract}


\section{Introduction}

In the recent decades, as a typical regional economic system, urban systems have attracted much attention from researchers. Series previous empirical studies has exhibited the basic feature of real-world urban systems: i) strong hierarchical structure, which usually means a few large cities are scattered among a large number of small and medium-sized cities, and the power-law-like city size distributions \cite{GX'99,Rozenfeld'11}. ii) urban agglomerations, in which serveral neighboring cities in some areas converge to integrate large metropolitan areas \cite{FY'17,Patricia'09}. iii) Allometric scaling relative growth, namely there are superlinear growth on urban economic output and sublinear growth on resource consumption \cite{LMA'07, Chen10}, and the economic output rate and resource utilization rate of large cities are relatively higher.
%The spatial structure of urban system has strong hierarchy, aggregation and allometry. City size distribution tends to approximate power-law form . , and .
In order to effectively describe and explain these properties of urban system, researchers have conducted a series of studies from many perspectives \cite{Dacey79,Bettencourt13}.
In this respect, Central Place Theory which was originally proposed by W. Christaller is one of the representative theoretical frameworks \cite{Christaller'66,Mulligan'12}. Its mainly assumptions are based on regional economics, that is, the city is based on the ``center" of providing various services. In an ideal uniform space, a stable central distribution based on hexagons will appear \cite{HSU'12,Taylor'10,Openshaw'03}. According to this theory, considering the different types of industral/commercial services, the resource requirements are different, and the services required for higher resources are gathered in large cities. This theroy has achieved a great success in explaining of the size distribution and industrial spatial distribution of cities \cite{Taylor'10,Openshaw'03}. In addition, the structural stability of urban aggregations and the identification of urban agglomerations in actual urban systems are also analyzed according to this theoretical framework \cite{FANG'07,Jordan'13}, showing the wide applicability of the theory.

In this paper, considering the basic model of Central Place Theory, we propose an novel method for the analysis of urban spatial distribution and cities' economic relationships based on the conversion of triangle mapping. Using this method, our empirical analysis of the spatial distribution of Chinese cities shows that the spatial location relationship between cities is sharply different in the cases of both economic volume heterogeneous and homogeneous. The difference shows that there are competitive and cooperative relationship between cities, and the two types of relationship change along the variation on the relative strength of economic volume of the two cities.

\section{Dataset}

We collect the data of urban economy of Chinese cities from the ``China Statistical Yearbook for Regional Economy" in the year of 2003, 2008 and 2013. The dataset includes the Gross Domestic Product (GDP) and the employment population for almost all the cities of mainland China in 2002, 2007 and 2012. Among them, the dataset of 2002, 2007 and 2012 contains information of 336, 337 and 337 cities respectively. At the same time, we get the latitude and longitude of each city's center position from the page of each city on Wikipedia (wikipedia.com).

\section{Empirical analysis}
\subsection{The nearest neighbor mapping triangle}

We analyze the geographical relations between cities by mapping the real-world position of cities to a regular triangle. The basic method of this mapping is that, for an observed city, we firstly define a regular triangle and put the transformed position of the observed city on the center of the regular triangle, and then transform the positions of the three nearest neighboring cities of the observed city onto the three corners of the regular triangle according to their geographical position and in the direction determined by a certain sequence of cities. The reason why we map to the regular triangle is that, on the one hand, the regular triangle structure is the subsystem of the hexagonal structure in the ideal model of Central Place Theory\cite{Christaller'66,Mulligan'12}; On the other hand, the triangles formed by the real-world positions of three neighboring cities must be convex, while the quadrilateral and other polygon transformations cannot ensure that the convexity of the polygon before and after the transformation is invariable. In addition, triangle transform minimizes the requirement of information of nearest neighboring cities and ensures the convenience of analysis.

Using the regular triangle transform, we first analyze the location relationship between each city and its three nearest neighboring cities. In this transform, the three nearest cities of the observed city are projected onto the three vertices ($0$, $\frac{\sqrt{3}}{3}$), ($-\frac{1}{2}$, $-\frac{\sqrt{3}}{6}$), ($\frac{1}{2}$, $-\frac{\sqrt{3}}{6}$) of the regular triangle with a length of 1 in order of GDP from high to low. At this point, the coordinates of the three nearest neighbor cities $\boldsymbol{A}_{i}(i = 1, 2, 3)$ and the mapping positions in the regular triangle $\boldsymbol{A^{'}}_{i}$ satisfy the following mapping relation:
\begin{equation}
%\begin{center}
\boldsymbol{A} \pmb{M} = \boldsymbol{A^{'}}_{i},
%\end{center}
\end{equation}
where $\pmb{M}$ is the matrix of coordinate transformation. According to the coordinates of the three nearest neighbors, the matrix of coordinate transformation $\pmb{M}$ is calculated and the mapping position of the observed city in the regular triangle is obtained: $\boldsymbol{B^{'}} = \boldsymbol{B}\pmb{M}$, where $\boldsymbol{B}$ is the original geographical location of the observed city. In the following discussions, the resulting triangle of the transformation is called the mapping triangle.

\begin{figure}
%\begin{center}
{\includegraphics[height=8.2cm]{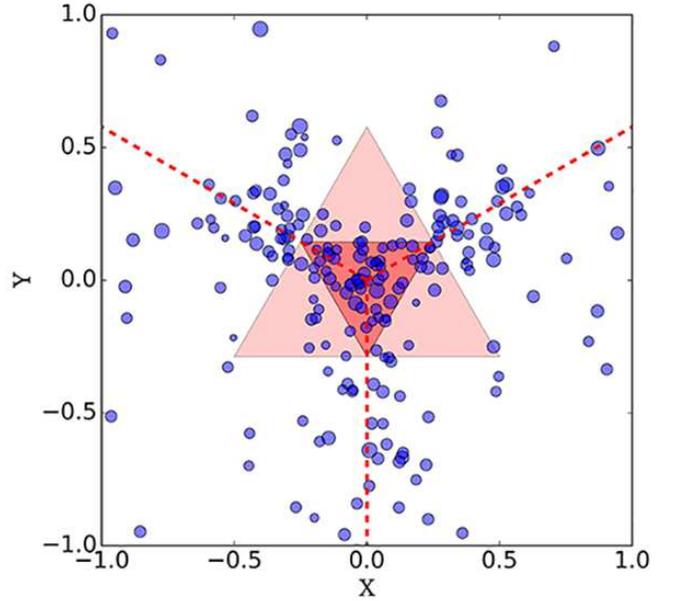}}
\caption{The relative position distribution of the cities in the mapping triangle (shallow red triangular area) of the three cities in its nearest neighbor in 2012. The three nearest neighboring cities are arranged counterclockwise from the vertex (0, $\sqrt{3}/3$) in descending order of GDP. The data point size is proportional to the logarithm of the city's GDP. The central triangle is shown by the deep red triangular region. The data point size is proportional to the logarithm of the city's GDP.}
%\end{center}
\label{Dissanjiaoxing2012}
\end{figure}

Fig. \ref{Dissanjiaoxing2012} shows the relative positional relationship of each city within the mapping triangle after the above triangle transform. It can be found that the relative position of each city is not evenly distributed in the mapping triangle, but concentrated near the center of the mapping triangle and the three side of the perpendicular direction. Since the three vertices of the mapping triangle are the projection positions of the three nearest neighbors of the city, this characteristic shows that the city tends to appear near the center of the gap in the neighboring cities, and the probability of near the three vertices is relatively low, suggesting that there is a spatial exclusion effect between the cities, that is, a relative position is too close the settlements of other cities are difficult to develop into new cities.

Considering that the data points are gathered near the center of the mapping triangles, we use two indicators to measure the centralization of data points: i) the proportion $q$ of cities in the central triangle of the mapping triangle (as shown in the deep red triangle region of the mapping triangle center of Fig. \ref{Dissanjiaoxing2012}); ii) the average distance $d$ of each city from the nearest midperpendicular of three sides in the mapping triangle. Since our analysis focus on the interaction between the city and its surrounding cities, we mainly investigate the data points that fall into the interior of the mapping triangle, which can ensure that the surrounding cities are non marginal and encircling. Therefore the calculation of the above two indicators is only for the data points inside the mapping triangle. In Fig. \ref{Dissanjiaoxing2012}, the total number of cities in the mapping triangle is $76$, in which the number of cities in the central triangle is $48$, and the $q$ is $0.63$. The average distance of each city from the nearest midperpendicular in the mapping triangle $d = 0.0409$.

We further construct a Null model to test the significance of the center-gathering tendency. In the Null model, we firstly randomly put $N$ points in a square space, and then calculate the mapping position of each point in mapping triangle that maps by its three nearest neighboring points using the same mapping method, and calculate the central triangle proportion $q_{0}$ and the average distance $d_{0}$ of the nearest midperpendicular. In the case of $N = 337$ (the number of cities in 2012), we numerically run the Null model with $10^{4}$ independent times and obtain $q_{0}=0.45$ and $d_{0}= 0.049$. Comparing with the results in Fig. \ref{Dissanjiaoxing2012}, we get the corresponding extreme probability $P_{q}(q_0 > q) = 0.001$ and $P_{d}(d_{0}< d) = 0.026$, indicating that the center-gathering tendency is significant, as shown in Fig. \ref{Dissanjiaoxing2012}.

\subsection{The mapping triangle based on GDP relationship}

In order to understand the geographical structure of the regional economy, we analyze the economic aggregate distribution of each city by using the mapping triangle method. We use GDP of each city as a measure of the total amount of the economy. Here the mapping method is that, we select the three nearest cities from all cities where the GDP is higher than the observed city itself, to map to the three corners of the triangle in counterclockwise order along the descending order of GDP (mapping the cite with highest GDP at position ($0$, $\frac{\sqrt{3}}{3}$), and then calculate the relative position of the observed city in the mapping triangle.
The reason why we use the nearest cities with higher GDP is that these cities would have stronger impact on the observed city.
%taking into account these near-scale and large-scale cities, the observed city may have stronger impact of the exclusive.

\begin{figure}
\centering
\subfigure
{\includegraphics[height=8.2cm]{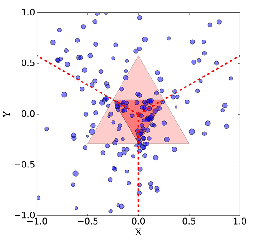}}
\subfigure
{\includegraphics[height=8.2cm]{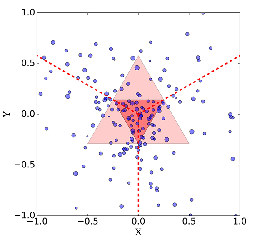}}
\caption{The distribution of each city in mapping triangle (the region in light pink triangle) for 2012 (the top panel) and 2002 (the below panel). They are composed by three nearest neighboring cities whose GDP is higher than that of the city. Three neighboring cities are arranged counterclockwise from the vertex (0, $\sqrt{3}$/3) in descending order of GDP. The central triangle is shown by the deep red triangular region. The data point size is proportional to the logarithm of the city's GDP.}
\label{GDPsanjiaoxing}
\end{figure}

Using this method, the locations of each city on mapping triangle based on the GDP are shown in Fig. \ref{GDPsanjiaoxing}. We can find that there is similar center-gathering property on the mapping triangle.
In order to test the significance of the center-gathering property, we also construct a Null model. Unlike the above Null model, due to this mapping method considers GDP of cities, this Null model aims to test the effect of GDP difference on real-world city positions. Therefore the original position of each city has become the background of statistical information and the city's GDP distribution will affect the statistical results, and we thus build the Null model by randomly exchanging the GDP of each pair of cities. In this Null model, each city's position is fixed as its real-world coordinate, but its GDP value is completely randomized by random exchange between different cities. TABLE \ref{Statistical parameters} lists the central triangle proportion $q$, the average distance $d$ of midperpendicular in 2012, 2007 and 2002, and the corresponding extreme case probability $P_{q}(q_{0} > q)$ and $P_{d}(d_{0} < d)$ of Null models for each year. However, comparing with the nearest neighbor mapping triangle based on only geographical location of cities, the significance of center-gathering property is obviously weak, and even its $d$ value does not show any obvious tendency, as shown in TABLE \ref{Statistical parameters}.

\begin{table}[htbp]
\caption{Results of mapping triangles based on GDP relation in each year.}
\label{Statistical parameters}\centering
\begin{tabular}{cccccc}
\hline
\hline
Year & $q$ & $P_{q}(q_{0} > q)$  & $d$ & $P_{d}(d_{0} < d)$ \\
\hline
2002 & 0.52 & 0.467 & 0.048 & 0.502  \\
2007 & 0.53 & 0.413 & 0.048 & 0.617  \\
2012 & 0.60 & 0.074 & 0.043 & 0.256  \\
\hline
\hline
\end{tabular}
\end{table}

\subsection{The transition between competition and cooperation} % of competition and exclusion between neighboring cities}

The construction of mapping triangles in the above two cases can be regarded as two special cases with different thresholds, and the threshold is the ratio of GDP between each neighboring city and the observed city (in the following discussions, the threshold is represented by $\mu$). In other words, the threshold $\mu$ means that, three angles of the mapping triangle correspond to the three nearest neighboring cities whose GDP reaches $\mu$ times of the observed city's GDP. The nearest neighbor mapping triangle therefore actually is same to the case with the threshold $\mu = 0$, and the mapping triangle based on the GDP relationship corresponds to the threshold $\mu = 1.0$.

%\ref{Proportion_ProbeSet} shows the improvements of SMD over MD and Hybrid methods with different size of the probe set.

\begin{figure}
\centering
%\begin{tabular}{ccc}
\includegraphics[width=8.8cm]{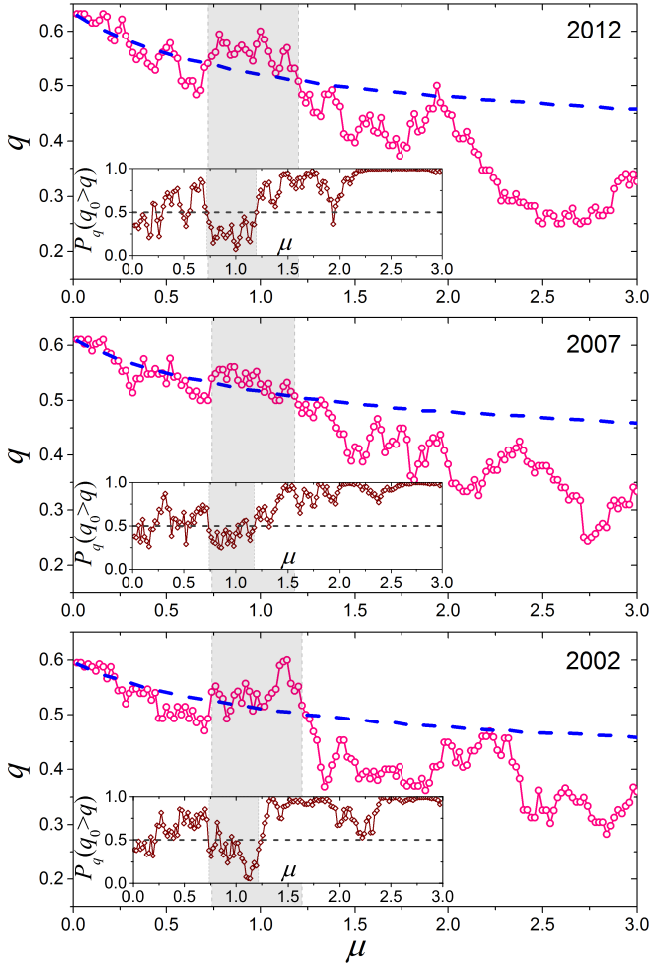} \\
%\end{tabular}
\caption{The central triangle proportion $q$ of each year along the change of threshold $\mu$ (The curve with pink data points in each panel). The blue dash line in each panel is the center triangle ratio $q_{0}$ along the change of $\mu$ in the Null model; The inset panels show the extreme case probability $P_{q}(q_{0} > q)$ vs. $\mu$ for the Null models of each year, where the gray dotted line represents $P_{q}(q_{0} > q) = 0.5$. The gray area indicates the window with center-gathering property, that is, the region of $\mu$ where $q_{0}$ is higher than the Null model predictive value.}
\label{threshold_reject}
\end{figure}

\begin{figure}
\centering
%\begin{minipage}[b]{0.5\textwidth}
\includegraphics[height=8.2cm]{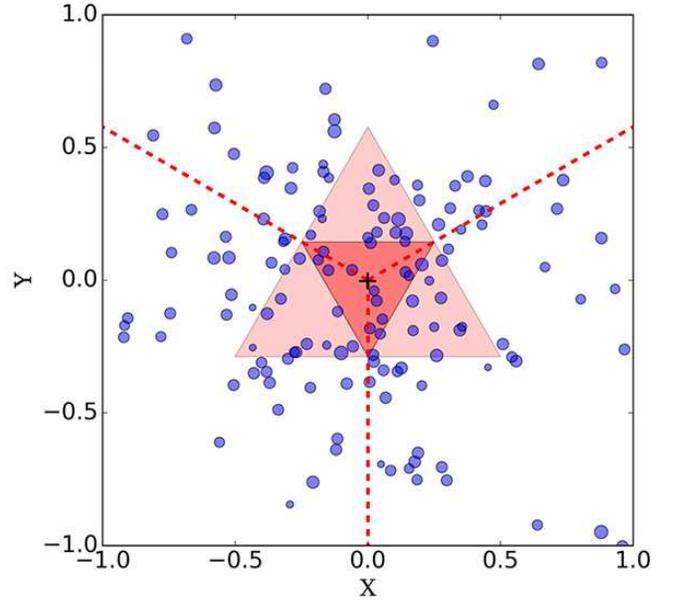}
%\includegraphics[totalheight=2in,bb=0 0 3408 3197]{GDPsanjiaoxing2002.jpg}
%\end{minipage}

%\includegraphics[width=0.225\textwidth]{dissanjiaoxing2012.eps} \\
%(a) Improvement of SMD&(b) Optimal parameters\\
\caption{The mapping triangle of the case $\mu$ = 2.5. Three neighboring cities are arranged counterclockwise from the vertex (0, $\sqrt{3}$/3) in descending order of GDP. The central triangle is shown by the deep red triangular region. The data point size is proportional to the logarithm of the city's GDP.}
\label{Yuzhi2.5GDPsanjiaoxing}

\end{figure}

Considering the significance of the center-gathering property obtained under $\mu = 0$ and $1.0$, it is necessary to explore the relationship between the center-gathering property and the threshold $\mu$. We calculated the central triangle proportion $q(\mu)$ for each year under different values of $\mu$. Furthermore, in the case of $\mu > 0$, the Null model is constructed by random exchange of GDP values of each pair of cities. The central triangle proportion $q_{0}(\mu)$ and the extreme case probability $P_{q}(q_{0} > q)(\mu)$ of the Null model are calculated. For a given $\mu$ value, if $q(\mu) > q_{0}(\mu)$, it indicates a possible center-gathering tendency; otherwise, the data points are more likely to be attracted to close the three vertices in the mapping triangle.

We find that the center-gathering property mainly emerges when $\mu$ is close to $1.0$, as indicated by the grey regions in Fig. \ref{threshold_reject}. The range of gray region in 2012, 2007 and 2002 respectively is $(0.72, 1.20)$, $(0.74, 1.18)$, $(0.74, 1.22)$, which is generally in the same. The result indicates that the economic relationship between neighboring cities with similar economic volume mainly is competition and exclusion. And the major in the region of $\mu$ outside the gray area did not show the center-gathering tendency. On the contrary, the property of corner gathering is more obvious. Especially in the case of $\mu > 1.5$,  $P_{q}(q_{0} > q) > 0.9$ for the major of this region, showing a relatively strong corner-gathering tendency, as the tyical case when $\mu = 2.5$ shown in Fig. \ref{Yuzhi2.5GDPsanjiaoxing}, implying that when a pair of neighboring cities have a large economic difference, the competitive exclusion relationship between them trends to weak and will be replaced by an attractive and cooperative relationship. According to the upper bounds of the gray regions in Fig. \ref{threshold_reject}, the critical point of the GDP ratio between the two cities from competition to cooperation is about $1.2$. In other words, the relationship between two neighboring cities will be change when their economic difference is higher than $1.2$ times.

%This phenomenon suggests that the economic relations between neighboring cities may change accordingly as the economic difference expands.

\section{Conclusion and discussion}

We propose a method that maps the position of each set of neighboring cities to a regular triangle projected from the relative position of its three neighboring cities. The distribution of the relative position of cities in the mapping triangle can reflect several features of the geographical and economic spatial structure of Chinese urban system. In fact, the triangle transition replaces the geographical location of cities by their relative position with neighboring cities to ensure the comparability of the spatial relationship for different cities.

Our analysis takes into account of the relationship of geographical location and the economic spatial distribution of urban system. Noticed that when urban economic effect is considered in our discussion, and the three corners of the mapping triangle usually are not the observed city's nearest cities. When the threshold $\mu$ is higher, the three corners of the mapping triangle correspond to the spatial position of the city may be far apart. In this case, the statistical patterns of the mapping triangle indicate the existence of long-range impact in urban system, especially for large cities.

Using this method, the observed features from urban systems include the following points: i) between the city and its nearest neighbors, obvious central agglomeration appears on the mapping triangle, suggesting that the nearest neighbor cities are dominated by competition and repulsion; ii) there is a obvious center-gathering effect for the relationship between cities and its nearest cities with close economic scale, implying stronger competition, while cities with large differences in the size of the neighboring cities show a triangular separation in the mapping triangle, revealing a strong trend of cooperation; iii) the tendency from competitive exclusion change to co-attraction, the corresponding urban GDP ratio of the threshold is about $1.2$. The transition of the relationship and the detailed process that is relevent to the underlying driven dynamics (e.g. the adjustment of its industrial structure and layout), and its role in the emergence of urban agglomerations, need deeper studies to investigate.

Furthermore, competition and cooperation between individuals often coexists in many types of socio-economic complex systems \cite{Turok04, Kapoor13}. The transition of relationship on the threshold that depends on the ratio of two neighboring cities' economic volume observed in urban system may also exists in other types of competing-cooperating systems, for example, the competition and cooperation between individuals, social groups or enterprises. This problem still need further empirical studies on these systems. In summary, the method of mapping triangle provides a novel insight in digging of spatial distributions and the tendency of relationships between spatial elements in urban system or other spatial systems.

%method of triangulation has a considerable applicability to discuss the spatial distribution characteristics of individuals in space systems where there is competition and cooperation mechanism at the same time.

%%%%%%------------------figure--ff_GlobalRankMD--------------------------%%%%%%
%\begin{figure}[tbh]
%%\centering
%\includegraphics[scale = 0.6]{ff_ep_GlobalRankMD.eps}
%\caption{The improvement of  recommendation results obtained by SMD comparing with GRM for new coming users. The length of recommendation list is $L=20$. For the sake of illustration, we only present the fractional improvement which is less than $100\%$ by the bar chart, and the values of the fractional improvements which are larger than $100\%$ are labeled on the corresponding bars.}
%\label{ff_GlobalRankMD}
%\end{figure}


\begin{thebibliography}{00}
\bibitem{GX'99} X. Gabaix. ``Zipf's law for cities: an explanation''. Quarterly Journal of Economics, vol. 114, no.3, pp. 739--767, 1999.
\bibitem{Rozenfeld'11} H. Rozenfeld, D. Rybski, X. Gabaix, H. Makse. ``The area and population of cities: new insights from a different perspective on cities''. American Economic Review, vol. 101, no. 5, pp. 2205-2225, 2011.
\bibitem{FY'17} C. Fang, D. Yu. ``Urban agglomeration: An evolving concept of an emerging phenomenon''. Landscape and Urban Planning, vol. 162, pp. 126-136, 2017.
\bibitem{Patricia'09} P.C. Melo, D.J. Graham, R.B. Noland. ``A meta-analysis of estimates of urban agglomeration economies''. Regional Science and Urban Economics, vol.39, pp. 332-342, 2009.
\bibitem{LMA'07} L.M.A. Bettencourt , J. Lobo, D. Helbing, C. K\"{u}hnert, G.B. West. ``Growth, innovation, scaling, and the pace of life in cities''. Proc. Natl. Acad. Sci. U.S.A. vol. 104, pp. 7301-7306, 2007.
\bibitem{Chen10} Y. Chen. ``Characterizing Growth and Form of Fractal Cities with Allometric Scaling Exponents''. Discrete Dynamics in Nature and Society, vol. 2010, pp. 194715, 2010.
\bibitem{Dacey79} M.F. Dacey. ``A Growth Process for Zipf's and Yule's City-Size Laws''. Environment and Planning A, vol. 11, no.4, pp. 361372, 1979.
\bibitem{Bettencourt13} L.M.A. Bettencourt. ``The Origins of Scaling in Cities''. Science, vol. 340, pp. 1438-1441, 2013.
\bibitem{Christaller'66} W. Christaller. $\textit{Central Places in Southern Germany}$. New Jersey: Prentice Hall, 1966.
\bibitem{Mulligan'12} G.F. Mulligan, M.D. Partridge, J.I. Carruthers. ``Central place theory and its reemergence in regional science''. Annals of Regional Science, vol. 48, pp. 405-431, 2012.
\bibitem{HSU'12} W.-T. Hsu. ``Central place theory and city size distribution''. The Economic Journal, vol. 122, pp. 903-932, 2012.
\bibitem{Taylor'10} P.J. Taylor, M.Hoyler, R. Verbruggen. ``External urban relational process: Introducing central flow theory to complement central place theory''. Urban Studies, vol. 47, pp. 2803-2818, 2010.
\bibitem{Openshaw'03}  S. Openshaw, Y. Veneris. ``Numerical experiments with central place theory and spatial interaction modelling''. Environment and Planning A, vol. 35, no. 8, 1389-1403, 2003.
\bibitem{FANG'07} C. Fang C, J. Song, D. Song. ``Stability of spatial structure of urban agglomeration in China based on central place theory''. Chinese Geographical Science, vol. 17 no. 3, pp. 193-202, 2007.
\bibitem{Jordan'13} J.W. Smith, M.F. Floyd. ``The urban growth machine, central place theory and access to open space''. City, Culture and Society, vol. 4, pp. 87-98, 2013.
\bibitem{Turok04} I. Turok. Cities, ``Regions and Competitiveness''. Regional Studies, vol. 38, pp. 1069-1083, 2004.
\bibitem{Kapoor13} R. Kapoor, J.M. Lee. ``Coordinating and competing in ecosystems: How organizational forms shape new technology investments''. Strategic Management J. vol. 34, no. 3, pp. 274-296, 2013.
\end{thebibliography}
\end{document}